\documentclass{article}
\usepackage{pgfplots}
\usepackage{booktabs} 
\usepackage{graphicx}
\usepackage{makeidx}
\usepackage{float}
\usepackage{caption}
\usepackage{booktabs}
\usepackage[backend=biber]{biblatex} 
\begin{document}

\title{Packing Peanuts: \\
The Role Synthetic Data Can Play in Enhancing Conventional Economic Prediction Models}

\author{Vansh Murad Kalia
\\
\\
Candidate for Master's of Arts in\\
Quantitative Methods for the Social Sciences
\\
\\
Columbia University
\\
\\
\\
Thesis Advisor: Prof. Gregory M. Eirich
\vspace{3cm}
}

\date{} % Leave this blank to remove the date

\maketitle

\begin{abstract}
\vspace{1cm}

    \noindent Packing peanuts, as defined by Wikipedia, is a common loose-fill packaging and cushioning material that helps prevent damage to fragile items. In this paper, I propose that synthetic data, akin to packing peanuts, can serve as a valuable asset for economic prediction models, enhancing their performance and robustness when integrated with real data. This hybrid approach proves particularly beneficial in scenarios where data is either missing or limited in availability. Through the utilization of Affinity credit card spending and Womply small business datasets, this study demonstrates the substantial performance improvements achieved by employing a hybrid data approach, surpassing the capabilities of traditional economic modeling techniques.

\end{abstract}

\clearpage
\section*{\centering Index}
\vspace{1cm}

\begin{table*}[h]
    \centering
    {\fontsize{12}{14}\selectfont 
    \renewcommand{\arraystretch}{2} 
    \begin{tabular}{|p{0.6\linewidth}|p{0.3\linewidth}|}
    \hline
    \textbf{Section} & \textbf{Page} \\
    \hline
    1. Introduction & 3 \\
    \hline
    2. Literature Review & 3 - 5 \\
    \hline
    3. Data & 5 \\
    \hline
    4. Methodology & 6 - 12 \\
    \hline
    \hspace{0.3cm} 4.1 Exploratory Data Analysis & 6 - 7 \\
    \hline
    \hspace{0.3cm} 4.2 Data Pre-processing & 8 \\
    \hline
    \hspace{0.3cm} 4.3 Model Selection & 8 - 10 \\
    \hline
    \hspace{0.3cm} 4.4 Model Testing Results & 10 - 12 \\
    \hline
    5. Conclusion & 12 \\
    \hline
    6. Limitations & 13 \\
    \hline
    7. Next Steps & 13 \\
    \hline
    References & 14 \\
    \hline
    \end{tabular}
    }
\end{table*}

\vspace{6cm}

\section{Introduction}

In recent years, the use of machine learning models for economic prediction has gained significant traction. While the adoption of these techniques has expedited the process of synthesizing vast amounts of data, one of the main challenges that remains is obtaining the data itself (or enough of it, at least!). Traditional approaches to data collection in the field of economics can be time-consuming, expensive, and limited in scope. There are countless cases where data is available but spotty, with missing samples. In such cases, synthetic data has emerged as a promising candidate to help fill that gap, but what is synthetic data?\\

On the highest level, synthetic data can be categorized into three main types:

\begin{itemize}
    \item Derived from real datasets, inheriting their statistical properties.
    \item Generated independently of real data,  without using any existing datasets.
    \item Hybrid in nature, combining aspects of the first two types.
\end{itemize}
This paper focuses on the Hybrid type, exploring its potential applications in enhancing economic prediction models.\\
\\
Utilizing data from Affinity and Womply, this study aims to investigate whether the integration of synthetic data can improve model performance and robustness in scenarios characterized by limited data availability, potentially outperforming models reliant solely on real data.

\section{Literature Review}

Given the nascent nature of the academic intersection of economic prediction models and synthetic data, there is not a lot of academic research that focuses directly on this topic. As such, for this research, I leverage some academic literature on synthetic data in relational fields like computer science, to formulate my hypothesis.
\\
\\
\textbf{Synthetic Data Generation for Economists:
}\footfullcite{koenecke2020synthetic}
\\
\\
In my search for academic literature at the intersection of synthetic data and economics, this paper stands out as one of the most important contemporary pieces. In this study, the authors address synthetic data generation within the field of economics by recognizing the challenges associated with accessing and handling sensitive or limited datasets. Koenecke and Varian discuss the methodologies and implications of generating synthetic data, providing economists with a valuable resource for exploring and testing hypotheses in situations where real data availability is constrained. The authors propose the use of synthetic data as an alternative for economic researchers:

\begin{itemize}
    \item Assist with privacy issues related to the use of data.
    \item Increase the number of samples available for a certain type of data.
    \item Test the robustness of existing models.
\end{itemize}

 \noindent The paper contributes as an important piece to my research by offering insights into the potential benefits of synthetic data and helping formulate my hypothesis that using the hybrid of synthetic and real data should improve the performance of an economic prediction model.
\\
\\
\textbf{Macroeconomic Predictions using Payments Data and Machine Learning:
}\footfullcite{chapman2022macroeconomic}
\\
\\
In this study, the authors focus on predicting the economy's short-term dynamics and delve into economic forecasting by leveraging payments data and machine learning techniques. This paper aims to demonstrate that non-traditional and timely data such as retail and wholesale payments, with the aid of nonlinear machine learning approaches, can provide policymakers with sophisticated models to accurately estimate key macroeconomic indicators in near real-time. By incorporating advanced machine learning algorithms and non-linear learning approaches, Chapman and Desai show over 40 percent improved accuracy in macroeconomic nowcasting. As a deeply quantitative study, this paper helped me structure the quantitative analysis for my research and nudged me towards the data I use as well.
\\
\\
\textbf{Augmentation Techniques in Time Series Domain: A Survey and Taxonomy}\footfullcite{Iglesias_2023}
\\
\\
This study offers a comprehensive overview of various data augmentation methods specifically tailored for time series data. In this paper, the authors delve into a systematic classification of different augmentation techniques, categorizing them based on their underlying principles and applications. The authors explore a wide array of augmentation approaches including traditional methods such as linear interpolation and synthetic data generation techniques like Generative Adversarial Networks (GANs). They discuss the advantages, limitations, and potential applications of each technique, providing insights into their effectiveness in addressing various challenges encountered in time series analysis. Additionally, the paper examines the implications of data augmentation on model generalization, robustness, and interpretability. Overall, this survey and taxonomy have helped me navigate the landscape of data augmentation techniques in the context of the time series analysis pertinent to my research.
\\
\\
\textbf{K-Nearest Neighbor (k-NN) based Missing Data Imputation}\footfullcite{inproceedings}
\\
\\
The authors of this paper explore the application of the K-Nearest Neighbor (k-NN) algorithm for imputing missing data. They investigate the use of the k-NN method as a means to address missing data in datasets and through their research, they propose a framework that leverages the k-NN algorithm to predict missing values based on the values of neighboring data points. This approach aims to improve data completeness and accuracy in datasets affected by missing information. While this paper contributes to the field of data imputation by offering a novel method that utilizes machine learning techniques to handle missing data effectively, it's not necessarily most suitable for my research as the distance between the missing data points is too much to be able to efficiently use the k-NN algorithm.

\section{Data}

For this research, I've opted to use the Affinity credit card spending datasets and Womply small business datasets from the Economic Tracker database\footnote{https://github.com/OpportunityInsights/EconomicTracker}. These datasets offer diverse features, but to narrow the focus for hypothesis testing, attention is given to the \texttt{daily\_spend\_19\_all} variable from Affinity and the \texttt{merchants\_all} variable from Womply. These variables can be described as follows:

\begin{itemize}
    \item \texttt{daily\_spend\_19\_all}: Daily spending in all merchant category codes (MCCs).
    \item \texttt{merchants\_all}: Percent change in the number of small businesses open, calculated as a seven-day moving average, seasonally adjusted, and indexed to January 4 to 31, 2020.
\end{itemize}

\noindent The \texttt{daily\_spend\_19\_all} variable comes from the Affinity dataset, as all spending features are measured relative to January 6 to February 2, 2020, seasonally adjusted, and calculated as a seven-day moving average. There are additional quartile features that are subdivisions by income using the median income of the ZIP codes; \texttt{q1} is the quartile with the lowest median income and \texttt{q4} is the quartile with the highest median income. I selected these variables on the highest level as they are ideal to test my hypothesis where there is missing data for the \texttt{merchants\_all} that I would look to impute.

\section{Methodology}

To test my hypothesis, I will create a real-life example using this dataset and my aim will be to create the best possible model to predict spending (\texttt{daily\_spend\_19\_all}) using the independent variable (\texttt{merchants\_all})

\subsection{Exploratory Data Analysis}

As an initial step in exploratory data analysis, I examine the descriptive statistics of the original dataset:
\begin{table}[htbp]
  \centering
  \caption{Descriptive Statistics}
    \begin{tabular}{lcc}
    \toprule
          & daily\_spend\_19\_all & merchants\_all \\
    \midrule
    count & 1253.000 & 109.000 \\
    mean  & 0.280 & -0.056 \\
    std   & 0.267 & 0.067 \\
    min   & -0.643 & -0.302 \\
    25\%  & 0.124 & -0.066 \\
    50\%  & 0.243 & -0.049 \\
    75\%  & 0.455 & -0.021 \\
    max   & 1.200 & 0.086 \\
    \bottomrule
    \end{tabular}%
  \label{tab:part1}%
\end{table}
\\
The descriptive statistics provide basic insights into the dataset. However, they do not offer much relevant information for addressing the research question. Thus, I proceed towards exploratory data analysis to examine the temporal distribution of the two variables.

\begin{figure}[H]
    \centering
    \includegraphics[width=0.8\linewidth]{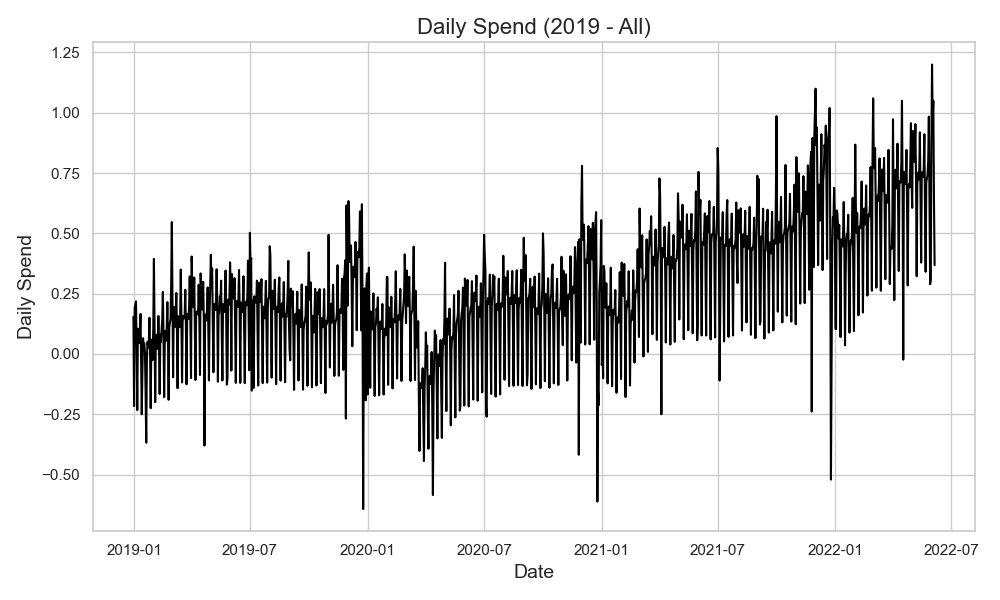}
    \caption{Data Distribution for Daily Spend}
    \label{fig:figure1-label}
\end{figure}
\begin{figure}[H]
    \centering
    \includegraphics[width=0.8\linewidth]{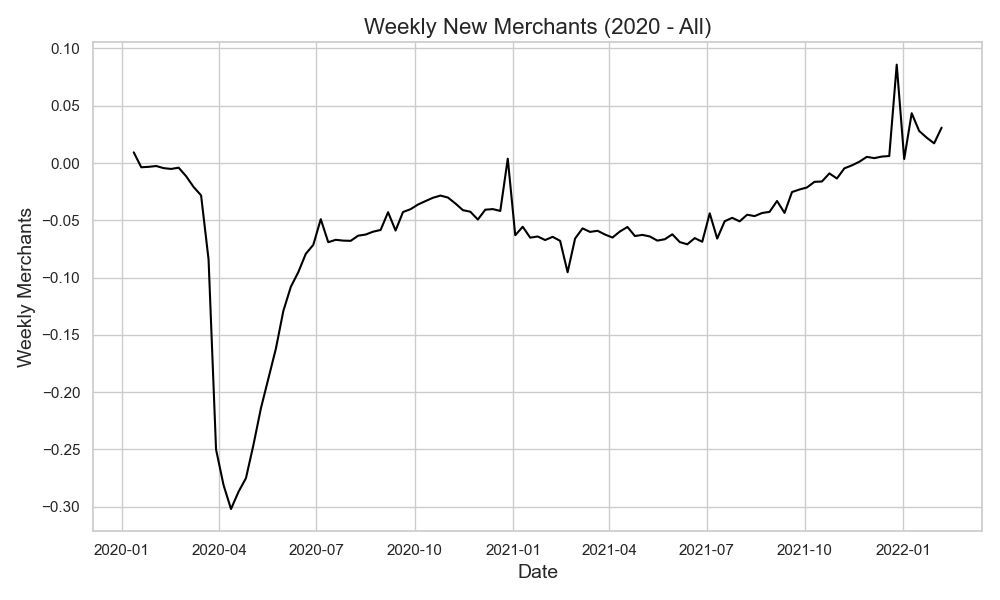}
    \caption{Data Distribution for All Merchants}
    \label{fig:figure2-label}
\end{figure}
\begin{figure}[H]
    \centering
    \includegraphics[width=1\linewidth]{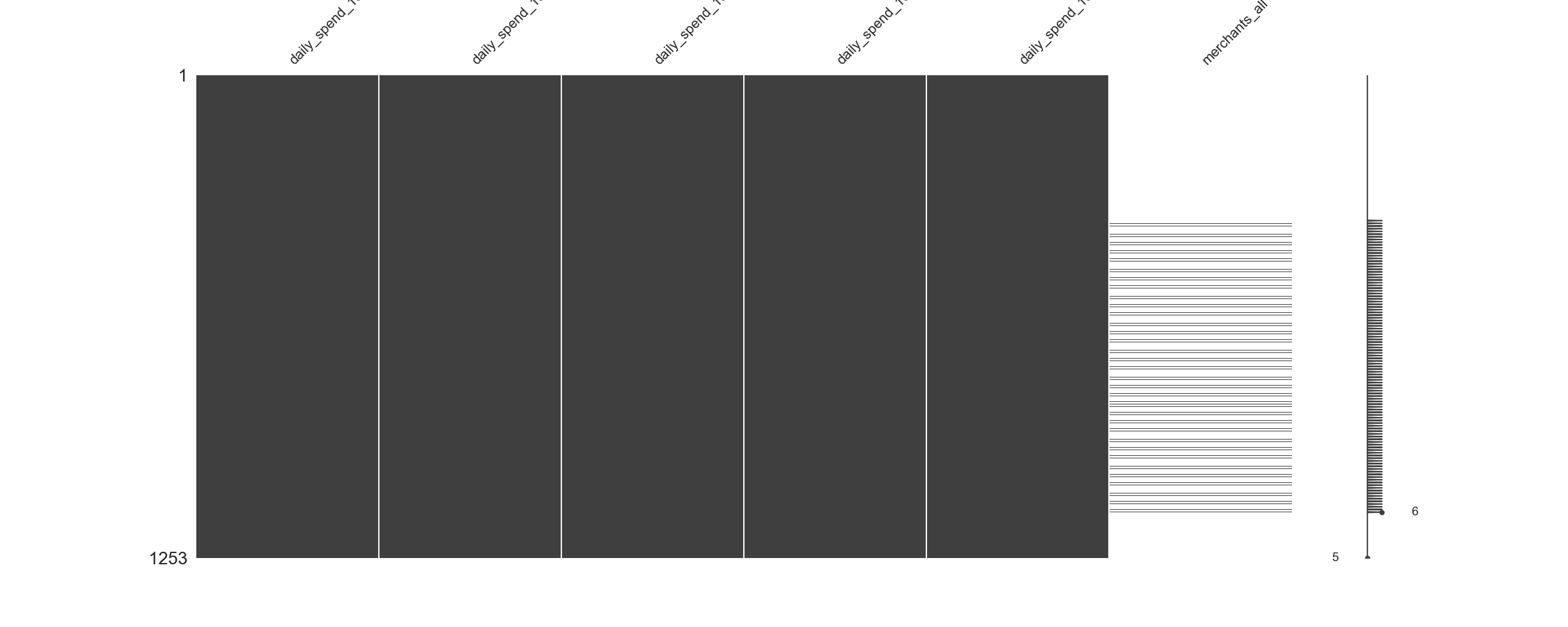}
    \caption{Missing Data Across Variables}
    \label{fig:figure3-label}
\end{figure}

\noindent Both Figure 1 and Figure 2 reveal notable trends, particularly a substantial drop in both variables during 2020, coinciding with the onset of the COVID-19 pandemic.  Additionally, there are intriguing outliers, such as the end of end-of-year data points in Figure 1. Figure 3 reveals that there are no missing variables for Daily Spend features but there are a lot of missing data for \texttt{merchants\_all}. Other than this, not much can be derived from an Exploratory Data Analysis that is relevant to my research question. The missing data will be addressed during preprocessing to ensure the success of this real-life example.   

\clearpage
\subsection{Data Pre-processing}
Despite the well-organized and clean nature of the data, several challenges exist, including datatype mismatches, DateTime information, filtering, and missing variables. To address these issues, I implement various data pre-processing techniques, such as handling different data types, managing DateTime information, and applying filtering. However, due to the density of the data, a careful selection of features is necessary for visualization purposes. Most notably, Womply's business data is provided on a weekly basis, in contrast to the daily spending data. This missing data could potentially affect the accuracy of my model's predictions and to address this discrepancy, I plan to generate synthetic data to fill this gap and facilitate meaningful comparisons with non-synthetic data models.\\
\\
I create four base datasets that cover conventional methods of missing data imputation used in Economics which will then be compared with the fifth model trained on the hybrid dataset built from generated synthetic data and real data. The techniques I follow are removing missing rows, global mean imputation, and Monte Carlo simulations and the base models for testing will be generated on the below datasets:

 \begin{itemize}
    \item Original dataset with no imputations
    \item Original dataset with missing rows removed
    \item Mean-imputed dataset that fills missing values using global mean
    \item Monte Carlo simulations imputed dataset
\end{itemize}

\noindent As mentioned in the literature review, I also considered the k-Nearest Neighbors (k-NN) technique for another base model. However, it proved unsuitable for the data I'm dealing with due to the lack of neighboring data points for proper imputation. These base datasets will facilitate the creation of the first four base models for testing and evaluation.

\subsection{Model Selection}
 As I begin the model selection process for testing and evaluation, it is important to recognize that this research involves a variety of complications with using time-series economic data, including missing data for a variable of interest and endogeneity concerns. As such, choosing the right model and evaluation techniques is of immense importance.\\
 \\
\textbf{OLS Regression:
}\\
\\
In the initial stage of analysis, I utilize Ordinary Least Squares (OLS) regression to investigate the linear relationship between the variables of interest. OLS regression is a widely used statistical method for estimating the relationship between a dependent variable and one or more independent variables by minimizing the sum of the squared differences between observed and predicted values\footfullcite{Zdaniuk2014}. By fitting a linear regression model to the data, I aim to identify any significant linear associations and quantify the strength and direction of these relationships. Additionally, OLS regression provides insights into the relative importance of each independent variable in explaining the variation observed in the dependent variable. This initial analysis will help inform subsequent modeling approaches and provide valuable insights into the underlying factors influencing the target variable's behavior.\\
\\
\textbf{Random Forest Model:
}\\
\\
Beyond capturing linear relationships, I employ a Random Forest model as a secondary economic prediction model. Random Forest is an ensemble learning method that constructs multiple decision trees during training and outputs the class that is the mode of the classes (classification) or mean prediction (regression) of the individual trees\footfullcite{Tripp2023DiabetIA}. It is a powerful tool for capturing nonlinear relationships and has been widely applied in economic research for variable selection, forecasting, and causal inference\footfullcite{coulombe2020macroeconomy}. The Random Forest algorithm is well-suited for handling complex relationships and interactions within the data, providing a more comprehensive understanding of the factors influencing the outcome. As such, I also use a Random Forest Model to compare the model performance across all the datasets.\\
\\
\textbf{Synthetic Data Generation:}
\\
\\
To complete the Model Selection process, I will now outline the unique approach I take to generate synthetic data to fill data gaps within the Womply business dataset. I train a Random Forest Model on the second dataset mentioned in the pre-processing section (original dataset with missing rows removed) as it represents the cleanest form of real data. While the original dataset contains missing daily values for \texttt{merchants\_all} variable, it still has enough weekly values for me to be able to use it as a target variable to train the random forest model. Then, I use the trained model to predict (or impute) the missing values within the original dataset.
\\
\\
This approach leverages Affinity data's \texttt{daily\_spend\_19\_all} features as inputs and learns any non-linear relationships between the target variable and the features. By doing so, it enhances the accuracy of filling the gaps in the \texttt{merchants\_all} column, leading to a more robust model. This leads to the creation of the fifth dataset for comparison against all the base models, and any model trained on this hybrid dataset will be referred to as Model 5 from here onwards.
\\
\\
Below is a scatter plot of the distribution of this hybrid dataset:

\begin{figure}[H]
    \centering
    \includegraphics[width=0.8\linewidth]{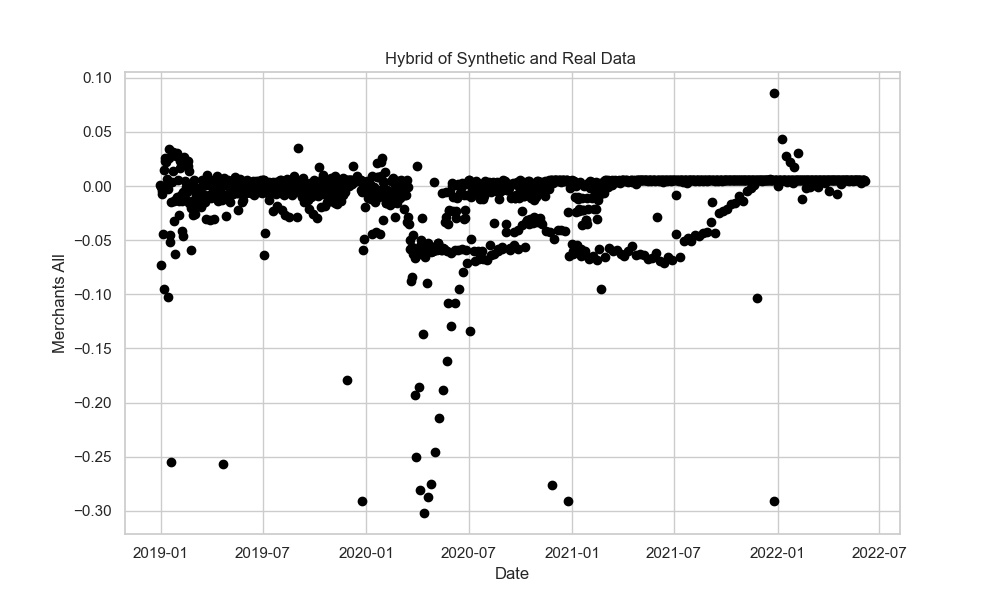}
    \caption{Hybrid Data Distribution for All Merchants}
    \label{fig:figure1-label}
\end{figure}

\noindent In comparison, below is a scatter plot of the distribution of just the real data:

\begin{figure}[H]
    \centering
    \includegraphics[width=0.8\linewidth]{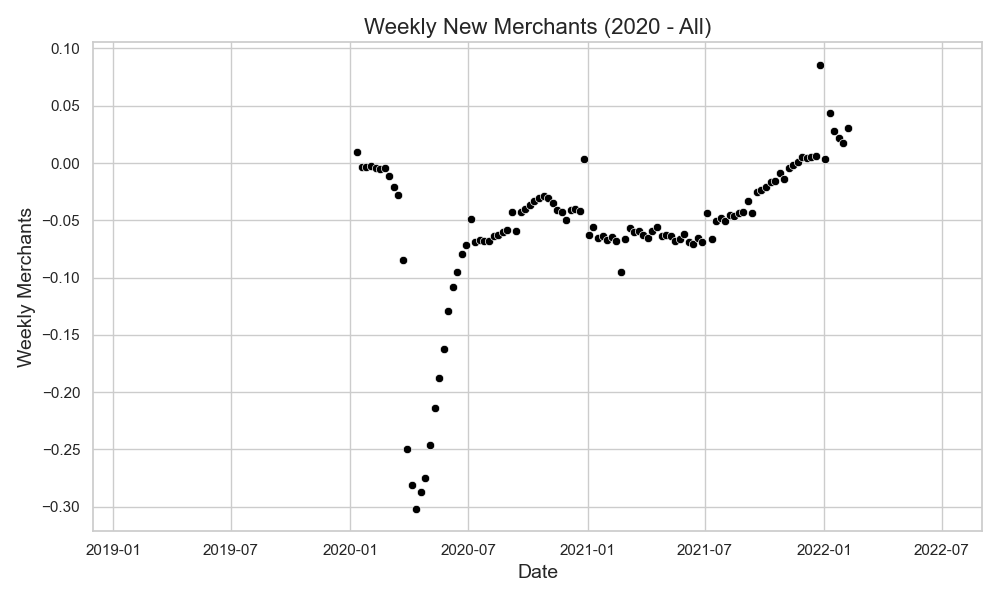}
    \caption{Hybrid Data Distribution for All Merchants}
    \label{fig:figure1-label}
\end{figure}

\clearpage
\subsection{Model Testing Results}
Now that all of the models are ready, let's take a look at how Model 5 (the comparison model) performs against all the base models across OLS Regression and Random Forest Models. 
\begin{center}
Base Models OLS Regression Results
\end{center}

\begin{center}
\begin{tabular}{lcccccc}
\toprule
\textbf{Model 1}          & \textbf{coef} & \textbf{std err} & \textbf{t} & \textbf{P$> |$t$|$} & \textbf{[0.025} & \textbf{0.975]}  \\
\midrule
\textbf{const}          &       0.0582***  &        0.017     &     3.369  &         0.001        &        0.024    &        0.092     \\
\textbf{merchants\_all} &       1.6710***  &        0.198     &     8.430  &         0.000        &        1.278    &        2.064     \\
\bottomrule
\end{tabular}
\end{center}\begin{center}
\begin{tabular}{lcccccc}
\toprule
\textbf{Model 2}          & \textbf{coef} & \textbf{std err} & \textbf{t} & \textbf{P$> |$t$|$} & \textbf{[0.025} & \textbf{0.975]}  \\
\midrule
\textbf{const}          &       0.0582***  &        0.017     &     3.369  &         0.001        &        0.024    &        0.092     \\
\textbf{merchants\_all} &       1.6710***  &        0.198     &     8.430  &         0.000        &        1.278    &        2.064     \\
\bottomrule
\end{tabular}
\end{center}\begin{center}
\begin{tabular}{lcccccc}
\toprule
\textbf{Model 3}         & \textbf{coef} & \textbf{std err} & \textbf{t} & \textbf{P$> |$t$|$} & \textbf{[0.025} & \textbf{0.975]}  \\
\midrule
\textbf{const}          &       0.3737***  &        0.023     &    16.473  &         0.000        &        0.329    &        0.418     \\
\textbf{merchants\_all} &       1.6710***  &        0.381     &     4.382  &         0.000        &        0.923    &        2.419     \\
\bottomrule
\end{tabular}
\end{center}\begin{center}
\begin{tabular}{lcccccc}
\toprule
\textbf{Model 4}          & \textbf{coef} & \textbf{std err} & \textbf{t} & \textbf{P$> |$t$|$} & \textbf{[0.025} & \textbf{0.975]}  \\
\midrule
\textbf{const}          &       0.0582***  &        0.017     &     3.369  &         0.001        &        0.024    &        0.092     \\
\textbf{merchants\_all} &       1.6710***  &        0.198     &     8.430  &         0.000        &        1.278    &        2.064     \\
\bottomrule
\end{tabular}
\end{center}\begin{center}
\begin{tabular}{lcccccc}
\toprule
\textbf{Model 5}          & \textbf{coef} & \textbf{std err} & \textbf{t} & \textbf{P$> |$t$|$} & \textbf{[0.025} & \textbf{0.975]}  \\
\midrule
\textbf{const}          &       0.3302***  &        0.006     &    51.379  &         0.000        &        0.318    &        0.343     \\
\textbf{merchants\_all} &       4.2133***  &        0.165     &    25.588  &         0.000        &        3.890    &        4.536     \\
\bottomrule
\end{tabular}
\end{center}

\noindent From the above table, it's clear that while all the models have statistically significant coefficients and constants, Model 5 stands out in comparison to the baseline models with its notably higher coefficient value (4.2133) for the variable \texttt{merchants\_all}, indicating a stronger impact on the dependent variable. Moreover, Model 5 exhibits lower standard errors for both the constant and \texttt{merchants\_all}, suggesting greater precision in the coefficient estimates. The high t-values (51.379 and 25.588 respectively) and extremely low p-values indicate high significance, further supporting the robustness of Model 5. Overall, Model 5 appears to offer a more accurate and statistically significant representation of the relationship between the variables compared to the other models. This result confirms my hypothesis, however, I will still look to substantiate it using Random Forest Models and see how Model 5 compares to the baseline models.\\
\\
\\

\begin{table}[h]
\centering
\caption{Random Forest Model Results}
\begin{tabular}{@{}llll@{}}
\toprule
\textbf{Model} & \textbf{Average MAE} & \textbf{Average MSE} & \textbf{Average R-squared} \\ \midrule
1                  & NA & NA & NA     \\
2                  & 0.162 & 0.042 & -5.92     \\
3                  & 0.217 & 0.077 & -0.75     \\
4                  & 0.232 & 0.088 & -1.06    \\
5                  & 0.092 & 0.017 & 0.55     \\ \bottomrule
\end{tabular}
\end{table}

\noindent Similarly to the OLS Regression results, Table 2 demonstrates Model 5's superior performance as it exhibits the lowest average Mean Absolute Error (MAE) of 0.092 and the lowest average Mean Squared Error (MSE) of 0.017, indicating the closest proximity of predicted values to the actual values compared to other models. Additionally, Model 5 achieves the highest average R-squared value of 0.55, suggesting that it explains a higher proportion of the variance in the dependent variable. It should be noted that Model 1 is marked NA as the dataset for Model 1 has missing values and is not suitable for a Random Forest analysis. Models 3 and 4 display poorer performance across all metrics while Model 2, although showing a low average MAE and MSE, has a substantially negative R-squared value, indicating poor model fit or potential overfitting. Therefore, Model 5 emerges as the most favorable choice among the presented Random Forest models, demonstrating superior predictive accuracy and model robustness.

\section{Conclusion}

I started this research to explore whether the integration of Synthetic Data could enhance model performance and robustness in scenarios characterized by limited data availability. Based on the literature review, I hypothesized that employing the hybrid approach of synthetic and real data should improve the performance of an economic prediction model, surpassing the efficacy of utilizing only real data. To test this hypothesis, I set up a real-life example using the Affinity and Womply datasets and created four different baseline models covering the conventional data-handling techniques used in economic prediction modeling. These techniques covered using the original dataset with no imputations, the original dataset removing rows with missing data, imputing missing values with global mean, and Monte Carlo simulations. My comparison model was trained on a dataset created using an advanced data augmentation technique that leverages Random Forest Models to generate Synthetic Data and use it in conjunction with real data. The comparison model outperformed all the baseline models across both OLS Regression testing and Random Forest Modeling, giving me strong conviction that my hypothesis is correct. 

\section{Limitations}

In terms of limitations of this paper, there were quite a few that I faced throughout the course of the research:

\begin{itemize}
    \item As highlighted in the literature review, the nascent nature of this topic meant that there was not a lot of reliable academic research I could leverage that focused directly on the intersection of economics and synthetic data. I consider this a huge limitation, as a lot more literature on the topic could have changed the structure of my research.
    \item The dataset I have been working with had a very high number of missing values for the target variable which could've easily led to an imbalanced dataset, adding bias in the generated data. If I had access to more data, or data with more frequency, I potentially could've used other baseline modeling techniques like k-NN and seen different results.
    \item Another limitation would be lacking the skills required to build more sophisticated models like Generative Adversarial Networks (GANs) or Variational Auto-Encoders (VAEs). Based on the literature review, it is very likely that synthetic data generated through either of these models would perform better than synthetic data generated by a Random Forest Model.
   
\end{itemize}

\noindent Even considering these limitations, I believe this research is well grounded in both qualitative and quantitative logic proving valid grounds for my hypothesis to be correct. 

\section{Next Steps}

In terms of next steps, I would like to create three more hybrid datasets using the following techniques that I discovered through my literature review \footfullcite{Iglesias_2023} and include them as comparison models to the existing tests: 

\begin{itemize}
 \item A library like Datawig to leverage Deep Learning Neural Networks.
 \item Generative Adversarial Network (GAN) to generate more accurate synthetic data.
 \item Variational Auto-Encoder (VAE) to generate more accurate synthetic data by accounting for variance in time-series data.
\end{itemize}

\noindent All of these are more sophisticated modeling techniques that have the potential to produce more robust results compared to Random Forest Models and are something I can look forward to implementing in economic prediction models.

\clearpage
\printbibliography

\end{document}